\documentclass{kluwer} 
\def\note #1]{{\bf #1]}} 
\def\be{\begin{equation}}
\def\ee{\end{equation}} 
\def\bearr{\begin{eqnarray}} 
\def\eearr{\end{eqnarray}}
\def\barr{\begin{array}}

\def\aap{\it Astron. Astrophys. \rm}

\def\apj{\it Astrophys. J. \rm}

\input psfig.sty
\begin{document}
\begin{article}
\begin{opening}
\title{VARIATION OF SOLAR IRRADIANCE AND MODE FREQUENCIES DURING MAUNDER MINIMUM}
\author{A. \surname{Bhatnagar}, Kiran \surname{Jain}
and S. C. \surname{Tripathy}}  
\institute{Udaipur Solar Observatory, Physical Research Laboratory, Off Bari Road,
 Dewali, P.  B.  No.  198, Udaipur 313004, India\\
\email{arvind@prl.ernet.in; kjain@prl.ernet.in; sushant@prl.ernet.in}}
\runningtitle{ESTIMATE OF SOLAR IRRADIANCE DURING MAUNDER MINIMUM}
\runningauthor{A. Bhatnagar, Kiran Jain and S. C.  Tripathy}

\begin{abstract} 
Using the sunspot numbers reported during the Maunder minimum and the 
empirical relations between the mode frequencies and solar activity 
indices, the variations in the 
total solar irradiance and 10.7 
cm radio flux  for the period 1645 to 1715 is estimated.  We find that 
the total solar irradiance and radio flux during the Maunder minimum decreased by 
0.19\% and  52\% respectively, as compared to the values for solar cycle 22. 
\end{abstract}
\date{\today}
\end{opening}

\section{Introduction} 

The period of low solar activity  between 1645 to 1715 is commonly 
known as the Maunder minimum and has attracted attention of  
researchers as it put forward
many unexpected questions. During this period,  the sunspots were
infrequent and as a result it was assumed that the solar dynamo and the solar
wind were either weak or switched off. Anomalies in the surface differential rotation 
during the Maunder minimum \cite{ribes93} also point out the global changes
in the internal solar dynamo mechanism. 
 Measurements of $^{10}$Be concentration in the Dye 3 ice core \cite{beer98}
recently showed that the magnetic cycles persisted throughout the Maunder minimum, 
although the Sun's over all activity was drastically reduced. This period
is also associated with the so called ``little ice age'' in Europe 
and \inlinecite{eddy76} suggested that it might be due to a decrease in  
solar irradiance. Many authors have estimated the total solar irradiance
during the Maunder minimum and find a decrease in the range of 0.1\% at a time of relatively high 
activity down to 1\% at a time of no or low activity.	

In a different approach, we estimate the change in the total 
solar irradiance and 10.7 cm radio flux during the Maunder minimum. This is 
achieved by calculating the  {\it p}-mode frequencies 
using the historic record of sunspots and the empirical
relations derived by \inlinecite{jain99}.  Since these relations between  
mode frequencies and solar activity indices are shown to be independent of 
solar cycles, these can be reliably used to estimate variations in   
solar irradiance and radio flux.

\begin{figure}
 \psfig{file=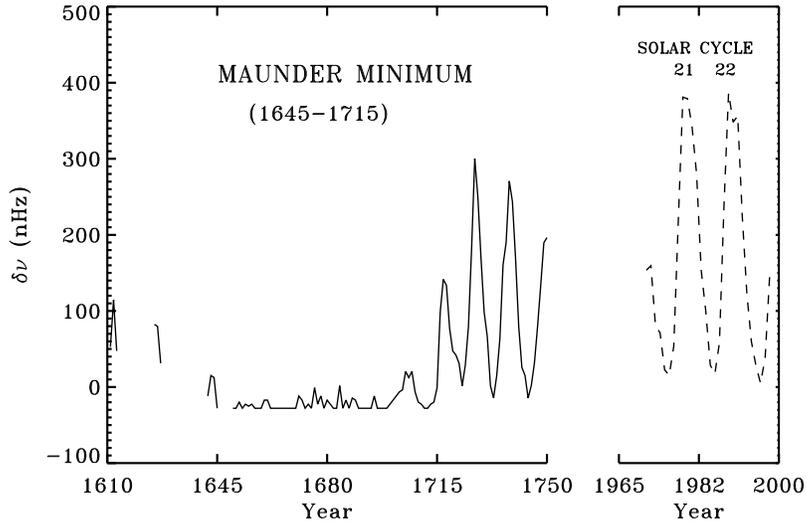,height=216pt}
\caption{Estimated frequency shift for the period 1610--1750 (solid line) and 1965--1998 (dashed line) 
with reference to the annual mean of 1996, using  the mean annual sunspot number 
from Eddy (1976) and Equation~(1).  
}  \label{fig1} 
\end{figure}  
\section{Estimates during Maunder Minimum} 
\subsection{{\it p}-Mode Frequencies}
We estimate the change in {\it p}-mode frequencies during the Maunder minimum  using
the following relation derived by \inlinecite{jain99}: 
  
\begin{eqnarray}
\delta\nu & =& (2.41 \pm 0.19)~\delta R_s - (0.48 \pm 1.68),
\end{eqnarray}
where $\delta \nu$ is given in nHz and  $R_s$ is the 
smoothed sunspot number. 
  	The estimated $\delta \nu$ for the epoch
  1610 -- 1750 and 1965 -- 1998 using the mean annual sunspot numbers from \inlinecite{eddy76}
  and Equation~(1) is plotted in Figure~1. It is observed that during the deep Maunder minimum
  (1655-1685), the frequency shifts are negligible corresponding to the low or no solar 
  activity period.   The maximum 
  change in the {\it p}-mode frequencies 
  during the entire period is  found to be 43 $\pm$ 3 nHz, which corresponds to 11\% of 
  the maximum change during the solar cycle 22. 
   This  small change 
  in frequency may be interpreted as due to the internal structural
  stability in the Sun during that period.
  
  	From the historic photoheliograpic records, \inlinecite{ribes88}  
   and \inlinecite{ribes93}  have shown that the solar
  diameter during the Maunder minimum increased by 7 arc seconds, as compared
  to the present value. \inlinecite{delache93}  have studied the correspondence of {\it p}-mode
  frequency shifts with solar radius and neutrino flux  and 
  found that $\delta \nu$ is anticorrelated with both.  From these findings,
  we infer that during the Maunder minimum, the neutrino flux would have increased
  from the present day value. This indicates that there was a real
  expansion of the solar envelope during this period and a decrease in solar 
  surface temperature and irradiance. These observed correlations between the solar activity,
  {\it p}-mode frequencies, solar radius, neutrino flux  and solar irradiance,
   strongly suggest the 
  possibility that the dynamical processes occuring in the upper layers are
  closely related to the perturbations inside the Sun, perhaps even down to the core. 
  
\subsection{Total Solar Irradiance and Radio Flux} 

   Since the Sun's radiation
is the principal driving force for global climate, the variation in total solar irradiance
(TSI) may lead to the climatic change. The coincidence of Maunder's ``prolonged 
solar minimum'' with the coldest period in Europe has been reported earlier.  It was suggested by \inlinecite{eddy76} that this cold
period might be due to  the reduction in  solar irradiance. The estimates for solar irradiance during
the Maunder minimum have been provided by many workers and they have found a decrease with respect to the
present value. To obtain an independent estimate of the variation of the solar irradiance,
we follow \inlinecite{jain99} and derive the following relation 
between $\delta\nu$ and $\delta TSI$:
\begin{eqnarray}
\delta\nu  = (68 \pm 16.4)~\delta TSI - (9.12 \pm 2.09)
 \end{eqnarray}
 	Using this relation  and the value of $\delta\nu$ for the
 Maunder minimum obtained earlier,  we estimate that the total solar irradiance 
 during Maunder minimum decreased
  by 2.68 W/m$^2$ (0.19\%) from the average value  of 1366.06 W/m$^2$
  for cycle 22. \inlinecite{reid91} had found that the TSI 
during Maunder minimum was lower
by 1\%  than the value for 1980. Using empirical correlation between TSI and the 
integrated Ca II emission, \inlinecite{lean92} suggested a decrease between 
0.15\% and 0.35\%.  
\inlinecite{nesme92} obtained a 0.2\% decrease in the irradiance by
comparing the variation in the 
differential helio-latitudinal rotation during the Maunder minimum and  present time. 
\inlinecite{nesme93} calculated an average decrease of 0.25\%--0.5\% in luminosity
 using the solar radius measurements. They had assumed an anticorrelation between the Sun's
diameter and its luminosity. \inlinecite{men96} modified these values to 0.11\%--0.43\% 
using the Maunder minimum sunspot numbers and near-equatorial rotation rates. Further, from the
contemporary and past rotation rates and solar radii, \inlinecite{men97}
reported a decrease of 1.23\% in TSI for the year 1683, when there were almost 
no sunspots and 0.37\% for the year 1715, when solar
activity begun to rise. Thus, our estimate of TSI fairly agrees with the earlier results 
using different acitivity indicators or physical parameters. 

	From the estimated change in frequency during the Maunder minimum
	and the following equation from \inlinecite{jain99} 
\begin{eqnarray}
\delta\nu  & = &(2.66 \pm 0.20)~\delta F_{10} - (4.67 \pm 1.44), 
\end{eqnarray}	
	 we calculate the change in 10.7 cm radio flux during Maunder 
minimum. As compared to 128 sfu - the average 
value for cycle 22, the radio flux is found to decrease by 66.2 sfu 
{\it i.e.} by 52\%  during the Maunder minimum. 

  \section{Conclusion} 
  
  	Using the  historical sunspot data and the  empirical relations
  	between the {\it p}-mode frequencies and activity indices 
derived by \inlinecite{jain99}, it is estimated that during the Maunder minimum
  from 1645 to 1715 AD, the mode frequencies showed a maximum change  of 
  43 $\pm$ 3 nHz. 
   Using the derived mode frequencies as a proxy of the solar activity 
  indicators, we have further estimated the
    total
  solar irradiance and 10.7 cm radio flux and find that these are lowered by 0.19\% 
and 52\%  
  respectively than the average values of the solar cycle 22. 
  The decrease in {\it p}-mode frequency and irradiance during 
  Maunder minimum probably indicates stability in the dynamical processes
  occuring in the solar convection zone and even deep down up to the core.   	

\acknowledgements
This work is 
partially supported under the CSIR Emeritus Scientist Scheme and Indo-US collaborative 
programme--NSF Grant INT-9710279.

\end{article}

\begin{thebibliography}{}
\bibitem[\protect\citeauthoryear{Beer, Tobias, and Weiss}{1998}]{beer98}Beer, J., Tobias, S.
and Weiss, N.: 1998, {\it Solar Phys.} {\bf 181}, 237.
\bibitem[\protect\citeauthoryear{Delache {\it et al.}}{1993}]{delache93}Delache, Ph., Gavryusev, V., Gavryuseva, E., Laclare, F.
 R\'{e}gulo, C. and Roca Cort\'{e}s, T.: 1993, \apj {\bf 407}, 801. 
\bibitem[\protect\citeauthoryear{Eddy}{1976}]{eddy76}Eddy, J. A.: 1976, {\it Science} {\bf 192}, 1189. 
\bibitem[\protect\citeauthoryear{Jain {\it et al.}}{2000}]{jain99}Jain, K., Tripathy, S. C.,  Bhatnagar, A. and Kumar, B.: 
2000, {\it Solar Phys.}, 192, 487.
\bibitem[\protect\citeauthoryear{Lean, Skumanich, and White}{1992}]{lean92}Lean, J., Skumanich, A. and White, O.R.: 1992: {\it Geophys. Res. Lett.} 
{\bf 19}, 1951.
\bibitem[\protect\citeauthoryear{Mendoza}{1996}]{men96} Mendoza, B.: 1996, {\it Geofis. Int.} {\bf 35}, 161.
\bibitem[\protect\citeauthoryear{Mendoza}{1997}]{men97} Mendoza, B.: 1997, \apj {\bf 483}, 523.
\bibitem[\protect\citeauthoryear{Nesme-Ribes and Mangeney}{1992}]{nesme92}Nesme-Ribes, E. and Mangeney, A.: 1992, {\it Radiocarbon} {\bf 34}, 263.
\bibitem[\protect\citeauthoryear{Nesme-Ribes {\it et al.}}{1993}]{nesme93}Nesme-Ribes, E., Ferreira, E. N., Sadourny, R., Le Trent, H. and
Li, Z. X.: 1993, {\it Geophys. Res.} {\bf 98}, 18923.
\bibitem[\protect\citeauthoryear{Reid}{1991}]{reid91}Reid, G.: 1991, {\it J. Geophys. Res.} {\bf 96}, 2835.
\bibitem[\protect\citeauthoryear{Ribes and Nesme-Ribes}{1993}]{ribes93}Ribes, J.C. and Nesme-Ribes, E.: 1993, \aap {\bf 276}, 549.
\bibitem[\protect\citeauthoryear{Ribes, Ribes, and Barrthalot}{1988}]{ribes88}Ribes, E., Ribes, J. C. and Barthalot R.: 1988, 
in J. Christensen-Dalsgaard and S. Frandsen (eds.) {\it Advance in Helio-
and Asteroseismology}, D. Reidel Publishing Company, 
Dodrecht, Holland, p. 227.
\end{thebibliography}
\end{document}